\documentstyle[12pt]{article}
\def\Was{W\c as}

\def\sstrut{$\strut\atop\strut$}
 
\begin{document}
\vsize=25.0 true cm
\hsize=16.0 true cm
\predisplaypenalty=0
\abovedisplayskip=3mm plus 6pt minus 4pt
\belowdisplayskip=3mm plus 6pt minus 4pt
\abovedisplayshortskip=0mm plus 6pt
\belowdisplayshortskip=2mm plus 6pt minus 4pt
\normalbaselineskip=14pt
\normalbaselines
\begin{titlepage}
 
\begin{flushright}
{\bf CERN-TH/99-119}
\end{flushright}
 
\vspace{0.5cm}
\begin{center}
{\bf\Large
The Monte Carlo Program KORALZ, for the
Lepton or Quark Pair Production at LEP/SLC Energies -- \\
 From version 4.0 to version 4.04
}\end{center}
 \vspace{0.3cm}
\begin{center}
  {\bf S. Jadach} \\
  {\em CERN, Theory Division, CH 1211 Geneva 23, Switzerland,\\ and}\\
   {\em Institute of Nuclear Physics,
        Krak\'ow, ul. Kawiory 26a, Poland}\\
\vspace{0.2cm}
   {\bf B.F.L. Ward} \\
   {\em Department of Physics and Astronomy,\\
   The University of Tennessee, Knoxville, TN 37996-1200, USA\\
   and \\
   SLAC, Stanford University, Stanford, CA 94309, USA}\\
and\\
\vspace{0.2cm}
   {\bf Z. W\c{a}s } \\
  {\em CERN, Theory Division, CH 1211 Geneva 23, Switzerland,\\ and}\\
   {\em Institute of Nuclear Physics,
        Krak\'ow, ul. Kawiory 26a, Poland}\\
\end{center}
 
\vspace{0.5cm}
\begin{center}
{\bf   ABSTRACT}
\end{center}
Brief information  on new features of KORALZ version 4.04 with respect
to version 4.0 is given.
The main difference is that the new version could be used at LEP2 energies,
i.e. up to 240 GeV centre-of-mass system energy.
The possibility to switch on different classes of anomalous couplings is also
included. 
\vskip 0.9 cm
\centerline{\it To be submitted to Computer  Physics Communications}
 \vspace{0.4cm}
\begin{flushleft}
{\bf 
 CERN-TH/99-119\\
May 1999}
\end{flushleft}
\footnoterule
\noindent
{\footnotesize
\begin{itemize}
\item[${\dagger}$]
Work supported in part by 
the US DoE contracts DE-FG05-91ER40627 and DE-AC03-76SF00515,
Polish Government grants 
KBN 2P03B08414, 
KBN 2P03B14715, 
Maria Sk\l{}odowska-Curie Joint Fund II PAA/DOE-97-316,
and Polish-French Collaboration within IN2P3.
\end{itemize}
}
 
\end{titlepage}

\noindent{\bf UPDATE SUMMARY}
\hfill\vskip 10pt
\noindent{\sl Title of the program:} KORALZ, version 4.04\ .
 
\noindent{\sl Reference to original program:}
Comput. Phys. Commun. {\bf 79} (1994) 503
 
\noindent{\sl Authors of original program:}
              S. Jadach, B.F.L. Ward and Z. W\c as
 
 
\noindent{\sl Operating system:}
UNIX, HP-UX 10.2
 
\noindent{\sl Programming language used:}
FORTRAN 77
 
\noindent{\sl High-speed storage required:}  $<$ 1 MB
 
\noindent{\sl No. of bits in a word:} 32
 
\noindent{\sl Peripherals used:} Line printer
 
 
 
\noindent{\sl Nature of the physical problem:}
Spin polarization of the $\tau$ in the process
$e^+e^-\rightarrow $ $\tau^+\tau^-(n \gamma ),$
                      $\tau^\pm\rightarrow X^\pm$
is used as an important data point for precise tests
of the standard electroweak theory. 
The  effects due to QED bremsstrahlung and apparatus efficiency have
to be subtracted from the data.
The program may be applied also to the production of  $u$, $d$,
$s$, $c$, $b$ quarks and neutrinos after appropriate redefinition
of the coupling constants and masses.
It may be used, as well, to simulate the muon pair production process.
The precision of the program in the case of quarks and electron neutrino 
is restricted.
Certain classes of anomalous couplings can be included for $\tau$, $q$
and $\nu$ final states. 
 
\noindent{\sl Method of solution:}
The Monte Carlo simulation of the combined $\tau$ production and decay
process is used to calculate the spin effects and effects of radiative
corrections, including hard bremsstrahlung, simultaneously.
Any experimental cut  and apparatus efficiency may easily be introduced
by rejecting some  of the generated events.
 
\noindent{\sl Restrictions on the complexity of the problem:}
The incoming $e^\pm$ and outgoing $\tau^\pm$ may have only
longitudinal  polarizations.
The total centre-of-mass energy is restricted  
to 240 GeV. The high precision of the program is assured
only in the region of the $Z$, where interference of bremsstrahlung
from initial and final states can be neglected.
 
 
\newpage
 
 \vskip 10pt
 
 The detailed description of how to use
the Monte Carlo program KORALZ can be found in ref. \cite{KORALZ}
and references therein, or at its WWW home page \cite{www}. 
 The KORALZ version 4.04 Monte Carlo can be used for simulation of  
$e^+e^- \to 2f\ n\gamma$, ($f=\mu,\tau,u,d,c,s,b, \nu$) processes,
including YFS exclusive exponentiation of initial-
and final-state bremstrahlung up to the LEP2 energy range. 
The differences with respect to the previously published version are not big.
However, as the LEP phase 2 is approaching its final years and its
collaborations will soon disappear, it is probably a good time to store 
and document, for future references and cross-checks, the final version 
of the KORALZ program. 
Let us note also that, in the future, KORALZ
will be replaced by the different program KK2f \cite{KK2f}, based on more
powerful exponentiation at the spin amplitude level. The
present paper is by no means an independent publication and it must be read as an 
appendix to  \cite{KORALZ}.

In the following, we give the minimum information on the content of the
present upgrade:

\vskip 1 cm
\centerline{\bf Standard Model Physics}
\vskip 1 cm

The  electroweak library DIZET version 5.0 \cite{DIZET} is used.
This code includes $ WW $  and $ ZZ $ boxes and other non-QED
corrections important at the LEP2 energy range. This library was
interfaced to KORALZ and the maximum allowed centre-of-mass energy
was shifted to 240 GeV.
The pretabulation of the electroweak form factors used by the generator
was extended to that range as well. It is of course straightforward to 
increase the maximum allowed energy to higher values, but only 
at the expense of a further increase of the initialization time and tests.
Other, less important modifications, minor bug fixes, etc.,  were also 
introduced into KORALZ and its libraries.

The basic algorithm was not modified and as precision requirements
at LEP2 energies are lower than at LEP1, where the program was
extensively tested and used, only a
limited set of tests were performed. Some of the numerical results of table
30  from ref. \cite{y-book} were reproduced, as well as results from
tables 1, 2, and 3 in the talk of  G. Passarino  \cite{passarino}. 
 
There are two technical
points that should be kept in mind. First, off the $ Z $ peak,
QED bremsstrahlung interference correction is no longer suppressed
below the  permille level. The size of this interference effect, not included in the
multiphotonic option of the program, can be checked by switching interference
corrections on and off \cite{AFBnowa} in single bremsstrahlung runs of KORALZ. 
Finally, let us point out that it is necessary,for high centre-of-mass energies,
to exclude from the generation the Born
singularity region (where the invariant mass of the $2f$ pair is
smaller than 2 GeV for example). This can be obtained by setting the
KORALZ input parameter XPR(13)=VVMAX to a value smaller than 1.

\newpage
 
\vskip 1 cm
\centerline{\bf Physics beyond the Standard Model}
\vskip 1 cm

In the course of experimental data analysis, the possibility to switch
on contributions from additional non-standard interactions on top of 
well controlled Standard Model predictions is of great
practical importance. This is an essential ingredient of any experimental 
program of the ``new physics'' searches. 
Some of such applications were developed on the basis of KORALZ and 
are included in the present version of the code.

In \cite{leptoki}  additional non-standard interactions 
introduced into KORALZ are due to 
the leptoquarks manifesting themselves in the channel 
$e^+e^- \to q \bar q (n\gamma's)$. A study of such an interaction was
of great interest in itself at some time. Now it can serve as
an example of how to implement any ``new physics'' interaction into the process
of fermion-pair production (with higher-order bremsstrahlung corrections).
In the table, we list the appropriate input parameters
for this particular non-standard interaction:

\vskip 1 cm
\vbox{
$$\halign{
\vrule  #
   & \sstrut {\tt #} \hfil \vrule
   & \hskip5pt \vtop{\hsize=11.5cm {\noindent \strut # \strut}}
   & # \vrule\cr
\noalign{\hrule}
& Parameter & Meaning          &\cr
\noalign{\hrule}
&NPR(13)=ifleptok    & Switch of leptoquarks 1/0 (on/off) &\cr
&NPR(14)=inter    &Type of the leptoquark interactions 
                   {\tt inter=}1, 2, $-$1,$-$2 corresponds respectively to 
                   the model 1, 2, 3, 4 of ref. \cite{leptoki} in case
                   of vector leptoquarks (or 3, 4, 1, 2 for scalar 
                   leptoquarks; this second option is however not directly
                   available to the user. The hard-coded flag {\tt iskal}
                   in function {\tt BORNDZ} has to be changed from 1 to 0
                   for that purpose).
                              &\cr
&XPR(19)=xmx & Mass of the leptoquark.&\cr
&XPR(20)=delta  &Additional parameter defining the size of the leptoquark
                 couplings.&\cr
&NPR(15)=ifkalin& Different, anomalous extensions 
                  of the Standard Model predictions,not discussed in this paper. 
                  The ifkalin$=$0 denotes
                  these contributions switched off. The ifkalin$=$1 denotes 
                  SM and anomalous $WW\gamma$ couplings for 
                  $e^+e^- \to \nu \bar \nu \gamma$ and ifkalin$=$2 anomalous 
                  gamma couplings for $e^+e^- \to \tau^+ \tau^- \gamma$. &\cr
\noalign{\hrule}
}$$}
\vskip 1cm
The input parameters are transmitted into the program with the help of the 
standard KORALZ
input matrices {\tt XPR}, {\tt NPR} (see \cite{KORALZ}).

Two additional interactions due to SM~\footnote{
At high energies the Standard Model $WW\gamma$ interaction becomes important
in the $e^+e^-\to \nu \bar \nu (n\gamma's)$ channel. The option {\tt ifkalin=1}
has to be used.
} and anomalous
$W-W-\gamma$ couplings in the channel $e^+e^-\to \nu \bar \nu (n\gamma's)$
and anomalous $\tau-\tau+\gamma$ couplings in the channel 
$e^+e^-\to \tau^+\tau^- (n\gamma's)$ are described separately in 
refs. \cite{Jachol} and \cite{Paul} respectively. The appropriate fortran code 
is included in the present program version, respectively
in directories {\tt korz{\_}new/nunulib} and 
{\tt korz{\_}new/ttglib}. The corresponding interactions are 
activated with the key 
{\tt NPR(15)} as explained in the table.

\vskip 1cm
\centerline{\bf Example demonstration programs}
\vskip 1cm
The demonstration program {\tt DEMO1.f} for the run when leptoquarks are
included can be found in the directory 
{ \tt korz{\_}new/february} and the output  {\tt DEMO1.out}\footnote{
The output {\tt DEMO1.out-no-leptoq}, with identical parameter settings
but leptoquarks switched off, can also be found there.}
 can be found
in thedirectory { \tt korz{\_}new/february/prod1}. The Standard Model {\tt DEMO.f}
and its output {\tt DEMO.out} are also included in these directories.

\vskip 1cm
\centerline{\bf Technical notes on KORALB and libraries used
by KORALZ}
\vskip 1cm
For the sake of completeness, let us mark that up-to-date versions of the $\tau$ decay library
TAUOLA \cite{TAUOLA} and KORALB program \cite{KORALB} for fermion 
pair production at less
than 30 GeV centre-of=mass energies are also included in the distribution file.
They differ from the respective published ones by minor bug-fixing only, 
flexibility extensions, etc. In the 
case of TAUOLA three new decay modes were introduced:
($\tau^+ \to K^+ K^0$, $\tau^+ \to 2\pi^+,  \pi^-, 3\pi^0$, and 
$\tau^+ \to 2\pi^+,  \pi^-, 3\pi^0$). But, as usual, no tuning of the decay 
parameters to the actual data is included.
The demonstration program of KORALB {\tt dist.f} can be found in the directory
{\tt korb/korb22/september} and its output {\tt dist.output} in 
{\tt korb/korb22/september/prod1}. For TAUOLA the corresponding files are
 respectively {\tt tauola/june/TAUDEMO.f}
and {\tt tauola/june/prod1/taudemo.out}.

An up to date version of PHOTOS \cite{PHOTOS} is included in the directory 
{\tt lib/photos.f} and its demonstration program {\tt phodem.f} in the directory
{\tt photos}. The differences with respect to the published version are mainly
technical. The present version is quite easy to adopt to the varying dimensions of
the HEPEVT \cite{HEPEVT} common block, and its internal variables are in double 
precision. A special routine devoted to the reparation of a 
kinematical configuration, affected, for instance, by the rounding errors,
 is also provided\footnote{ It is necessary
in case of PHOTOS operation on events including ultrarelativistic ($TeV$-range)
decaying light particles stored in the single precision variables.}
although it is not activated.
The interference correction in the case of two-body decays into particles
of different masses is now included.

The DIZET library for the calculation of genuine electroweak non-QED corrections
\cite{DIZET}, version 5.0, is placed in the directory  {\tt korz{\_}new/diz4{\_}9} 
(the older version 4.9 is also placed there) 
together with the interface.
An even older version can be found in the directory {\tt korz{\_}new/diz4{\_}6};  
the electroweak library Z0POLE \cite{Z0POLE}, with its demonstration program,
output and documentation, is in the directory  {\tt korz{\_}new/z0pole}.
In the actual runs of our demonstration programs, DIZET version 5.0 is used,
but it can be rather easily replaced with any of the others. 
However, not all options of KORALZ operations will then be assured.
 
There are two versions of YFS3 multiphoton generator included in KORALZ.
The version 3.0 \cite{YFS3.0} was actually used in most of the KORALZ
applications until now. The version 3.4  is most up to date;
it has an improved treatment of the configurations with very hard photons.
It is thus an intermediate step toward KK2f \cite{KK2f}. The demonstration outputs
were produced with the YFS version 3.4, but were checked to work with version
3.0 as well.

The description of libraries  of anomalous $\gamma$ couplings
\cite{Jachol} and \cite{Paul} are up to date. The libaries are placed 
in directories: {\tt korz{\_}new/nunulib} and {\tt korz{\_}new/ttglib}.
 
\vskip 1cm
\centerline{\bf Acknowledgements}
\vskip 1cm
We thank the CERN Theory Division and all four LEP collaborations for 
their support.
One of us (BFLW) thanks Prof. C. Prescott for the hospitality of the SLAC group A
while this work was completed.
We would like to thank all authors of libraries used in 
the program for their cooperation and support. Finally thanks go to all 
users of the progam for their numerous remarks, which were essential in shaping
the program as well as in bug-fixing. Without this massive support the present
project would be impossible.


\end{document}